\documentclass[runningheads,a4paper]{llncs}

\usepackage{amssymb}
\usepackage{graphicx}
\usepackage{xspace}
\usepackage{amsmath}
\usepackage{amssymb}
\usepackage{booktabs}
\usepackage{algorithm}
\usepackage[noend]{algpseudocode}
\usepackage{mathptmx}
\usepackage{url}
\usepackage{color}
\usepackage{fancyhdr}
\renewcommand{\footrulewidth}{0.4pt}
\urldef{\mailsa}\path|{alfred.hofmann, ursula.barth, ingrid.haas, frank.holzwarth,|
	\urldef{\mailsb}\path|anna.kramer, leonie.kunz, christine.reiss, nicole.sator,|
	\urldef{\mailsc}\path|erika.siebert-cole, peter.strasser, lncs}@springer.com|    
\newcommand{\keywords}[1]{\par\addvspace\baselineskip
	\noindent\keywordname\enspace\ignorespaces#1}

\newcommand{\mflash}{M-Flash\xspace}

\begin{document}
	\sloppy
	\mainmatter
	\title{\mflash: Fast Billion-scale Graph Computation Using a Bimodal Block Processing Model}
	
	\titlerunning{\mflash}
	
	\author{
		Hugo Gualdron\textsuperscript{1}, Robson Cordeiro\textsuperscript{1}, Jose Rodrigues-Jr\textsuperscript{1}, \\Duen Horng (Polo) Chau \textsuperscript{2}, Minsuk Kahng\textsuperscript{2}, U Kang\textsuperscript{3}
	}
	
	\authorrunning{\ }
			
	\institute{\textsuperscript{1}University of Sao Paulo, Sao Carlos, SP, Brazil\\
		\textsuperscript{2}Georgia Institute of Technology, Atlanta, USA\\
		\textsuperscript{3} Seoul National University, Republic of Korea\\
		\email{\{gualdron,robson,junio\}@icmc.usp.br, \{polo,kahng\}@gatech.edu, ukang@snu.ac.kr}}
	
	\maketitle
	
	\begin{abstract}
		Recent graph computation approaches have demonstrated that a single PC can perform efficiently on billion-scale graphs. While these approaches achieve scalability by optimizing I/O operations, they do not fully exploit the capabilities of modern hard drives and processors. To overcome their performance, in this work, we introduce the Bimodal Block Processing (\emph{BBP}), an innovation that is able to boost the graph computation by minimizing the I/O cost even further. With this strategy, we achieved the following contributions: (1) \mflash, the fastest graph computation framework to date; (2) a flexible and simple programming model to easily implement popular and essential graph algorithms, including the \textit{first} single-machine billion-scale eigensolver; and (3) extensive experiments on real graphs with up to 6.6 billion edges, demonstrating M-Flash's consistent and significant speedup.
	\end{abstract}

	\section{Introduction}
	Large graphs with {\it billions} of nodes and edges are increasingly common in many domains and applications, such as in studies of social networks, transportation route networks, citation networks, and many others.
	Distributed frameworks (find a thorough review in the work of Lu {\it et al.} \cite{lu2014large}) have become popular choices for analyzing these large graphs. 
	However, distributed approaches may not always be the best option, because they can be expensive to build \cite{Kyrola:2012:GLG:2387880.2387884}, and hard to maintain and optimize.
	
	These potential challenges prompted researchers to create single-machine, billion-scale graph computation frameworks that are well-suited to essential graph algorithms, such as eigensolver, PageRank, connected components and many others.
	Examples are GraphChi~\cite{Kyrola:2012:GLG:2387880.2387884} and TurboGraph~\cite{Han:2013:TFP:2487575.2487581}.
	Frameworks in this category define sophisticated processing schemes to overcome challenges induced by limited main memory and poor locality of memory access observed in many graph algorithms. However, when studying previous works, we noticed that despite their sophisticated schemes and novel programming models, they do not optimize for disk operations and data locality, which are the core of performance in graph processing frameworks.	
	
	In the context of {\it single-node}, {\it billion-scale}, graph processing frameworks, we present \textbf{\mflash}, a novel scalable framework that overcomes critical issues found in existing works. The innovation of \mflash confers it a performance many times faster than the state of the art. More specifically, our contributions include:
	\begin{enumerate}
		\item{{\bf \mflash Framework \& Methodology:} we propose a novel framework named \mflash that achieves fast and scalable graph computation. \mflash (\url{https://github.com/M-Flash}) introduces the Bimodal Block Processing, which significantly boosts computation speed and reduces disk accesses by dividing a graph and its node data into blocks (dense and sparse) to minimize the cost of I/O.}
		
		\item{\textbf{Programming Model:} \mflash provides a flexible and simple programming model, which supports popular and essential graph algorithms, e.g., PageRank, connected components, and the \textit{first} single-machine eigensolver over billion-node graphs, to name a few.}
		
		\item{{\bf Extensive Experimental Evaluation:} we compared \mflash with state-of-the-art frameworks using large graphs, the largest one having 6.6 billion edges (YahooWeb \url{https://webscope.sandbox.yahoo.com}). \mflash was consistently and significantly faster than GraphChi \cite{Kyrola:2012:GLG:2387880.2387884}, X-Stream \cite{Roy:2013:XEG:2517349.2522740}, TurboGraph \cite{Han:2013:TFP:2487575.2487581}, MMap \cite{LinKSCLK14}, and GridGraph \cite{190490} across all graph sizes. Furthermore, it sustained high speed even when memory was severely constrained, e.g., 6.4X faster than X-Stream, when using 4GB of Random Access Memory (RAM).}
	\end{enumerate}

	\section{Related Works}\label{sec:related}
	A typical approach to scalable graph processing is to develop a distributed framework. This is the case of Gbase \cite{kang2012gbase}, Powergraph, Pregel, and others \cite{lu2014large}. Among these approaches, Gbase is the most similar to \mflash. Despite the fact that Gbase and \mflash use a block model, Gbase is distributed and lacks an adaptive edge processing scheme to optimize its performance. Such scheme is the greatest innovation of M-Flash, conferring to it the highest performance among existing approaches, as demonstrated in Section \ref{sec:evaluation}.
	
	Among the existing works designed for single-node processing, some of them are restricted to SSDs. These works rely on the remarkable low-latency and improved I/O of SSDs compared to magnetic disks. 
	This is the case of TurboGraph \cite{Han:2013:TFP:2487575.2487581}, which relies on random accesses to the edges --- not well supported over magnetic disks. Our proposal, \mflash, avoids this drawback by avoiding random accesses.
	
	GraphChi \cite{Kyrola:2012:GLG:2387880.2387884} was one of the first single-node approaches to avoid random disk/edge accesses, improving the performance over mechanical disks. GraphChi partitions the graph on disk into units called \textit{shards}, requiring a preprocessing step to sort the data by source vertex. GraphChi uses a vertex-centric approach that requires a shard to fit entirely in memory, including both the vertices in the shard and all their edges (in and out). 
	As we demonstrate, this fact makes GraphChi less efficient when compared to our work. \mflash requires only a subset of the vertex data to be stored in memory.
	
	MMap \cite{LinKSCLK14} introduced an interesting approach based on OS-supported mapping of disk data into memory (virtual memory). It allows graph data to be accessed as if they were stored in unlimited memory, avoiding the need to manage data buffering. Our framework uses memory mapping when processing edge blocks but, with an improved engineering, \mflash consistently outperforms MMap, as we demonstrate. 
	
	GridGraph \cite{190490} divides the graphs into blocks and processes the edges reusing the vertices' values loaded in main memory (in-vertices and out-vertices). Furthermore, it uses a two-level hierarchical partitioning to increase the performance, dividing the blocks into small regions that fit in cache. When comparing GridGraph with \mflash, both divide the graph using a similar approach with a two-level hierarchical optimization to boost computation. However, \mflash adds a bimodal partition model over the block scheme to optimize even more the computation for sparse blocks in the graph.
	
	GraphTwist \cite{zhou2015graphtwist} introduces a 3D cube representation of the graph to add support for multigraphs. The cube representation divides the edges using three partitioning levels: slice, strip, and dice. These representations are equivalent to the block representation (2D) of GridGraph and \mflash, with the difference that it adds one more dimension (slice) to organize the edge metadata for multigraphs. The slice dimension filters the edges' metadata optimizing performance when not all the metadata is required for computation. Additionally, GraphTwist introduces pruning techniques to remove some slices and vertices that they do not consider relevant in the computation.
	
	\mflash also draws inspiration from the edge streaming approach introduced by X-Stream's processing model \cite{Roy:2013:XEG:2517349.2522740}\cite{Cheng2015}, improving it with fewer I/O operations for dense regions of the graph. Edge streaming is a sort of stream processing referring to unrestricted data flows over a bounded amount of buffering. As we demonstrate, this leads to optimized data transfer by means of less I/O and more processing per data transfer.
	
	\begin{figure}[!t]
		\centering
		\includegraphics[height=150px]{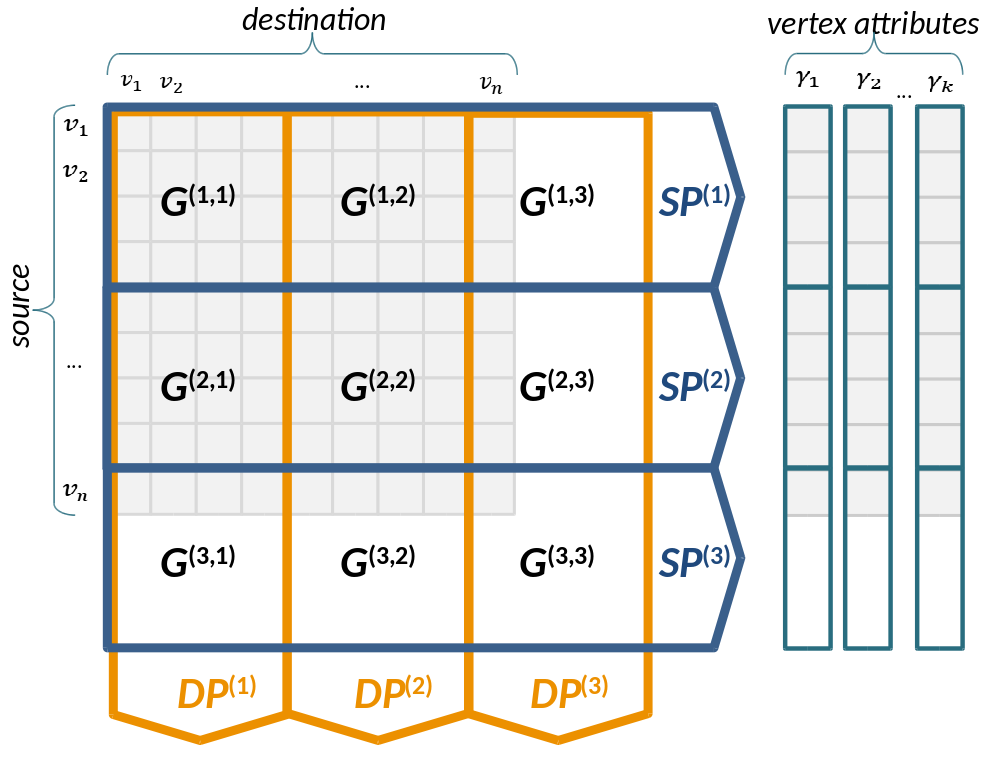}      
		\caption{
			Organization of edges and vertices in \mflash.
			\textbf{Edges (left)}: example of a graph's adjacency matrix (in light blue color) using 3 logical intervals ($\beta = 3$); $G^{(p,q)}$ is an edge block with source vertices in interval $I^{(p)}$ and destination vertices in interval $I^{(q)}$; $SP^{(p)}$ is a \emph{source-partition} containing all blocks with source vertices in interval $I^{(p)}$;$ DP^{(q)}$ is a \emph{destination-partition} containing all blocks with destination vertices in interval $I^{(q)}$.
			\textbf{Vertices (right):} the data of the vertices as $k$ vectors ($\gamma_1$ ... $\gamma_k$), each one divided into $\beta$ logical segments.
		}
		\label{fig:blocks}
	\end{figure}
	
	\section{\mflash}\label{sec:mflash}
	The design of \mflash considers the fact that real graphs have a varying density of edges; that is, a given graph contains dense regions with many more edges than other regions that are sparse. In the development of \mflash, and through experimentation with existing works, we noticed that these dense and sparse regions could not be processed in the same way. We also noticed that this was the reason why existing works failed to achieve superior performance. To cope with this issue, we designed \mflash to work according to two distinct processing schemes: Dense Block Processing (DBP) and Streaming Partition Processing (SPP). For full performance, \mflash uses a theoretical I/O cost-based scheme to decide the kind of processing to use in face of a given block, which can be dense or sparse. The final approach, which combines DBP and SPP, was named Bimodal Block Processing (BBP).
	
	\subsection{Graph Representation in \mflash}\label{section:representation}
	A graph in \mflash is a directed graph $G = (V,E)$ with vertices $v \in V$ labeled with integers from 1 to $\left | V \right |$, and edges $e = (source, destination)$, $e \in E$. Each vertex has a set of attributes $\gamma = \{\gamma_{1},\gamma_{2}, \dots, \gamma_{K} \}$; edges also might have attributes for specific processings.\\
	
	\noindent\textbf{\textit{Blocks} in \mflash}: 
	Given a graph $G$, we divide its vertices $V$ into $\beta$ intervals denoted by $I^{(p)}$, where $1 \leq p \leq \beta$. Note that $I^{(p)} \cap  I^{(p')} = \varnothing$ for $p \neq p'$, and $\bigcup _{p} \ I^{(p)} =V$. 
	Consequently, as shown in Figure \ref{fig:blocks}, the edges are divided into $\beta^2$ \textit{blocks}.
	Each block $G^{(p,q)}$ has a {\it source node interval} $p$ and a {\it destination node interval} $q$, where $1 \leq p,q \leq \beta$. 
	In Figure \ref{fig:blocks}, for example, $G^{(2,1)}$ is the block that contains edges with source vertices in the interval $I^{(2)}$ and destination vertices in the interval $I^{(1)}$. We call this on-disk organization as {\it partitioning}. Since \mflash works by alternating one entire block in memory for each running thread, the value of $\beta$ is automatically determined by the following equation:
	{\footnotesize
		\begin{equation}
			\beta = \left \lceil \frac{\phi (T + 1) \left | V \right |}{M} \right \rceil
			\label{eq:beta}
		\end{equation}}
		\noindent{where the constant $1$ refers to the need of one buffer to store the input vertex values that are shared between threads (read-only), $\phi$ is the amount of data to represent each vertex, $T$ is the number of threads, $\left | V \right |$ is the number of vertices, and $M$ is the available RAM. For example, 4 bytes of data per node, 2 threads, a graph with 2 billion nodes, and for 1 GB RAM, $\beta = \lceil (4 \times (2 + 1) \times (2\times10^9)) / (2^{30}) \rceil = 23$, thus requiring $23^2 = 529$ blocks.    The number of threads enters the equation because all the threads access the same block to avoid multiple seeks on disk, and they use an exclusive memory buffer to store the vertex data processed (one buffer per thread), so to prevent ``race'' conditions.} 
		
		\begin{figure}[!t]
			\centering
			\includegraphics[width=230px]{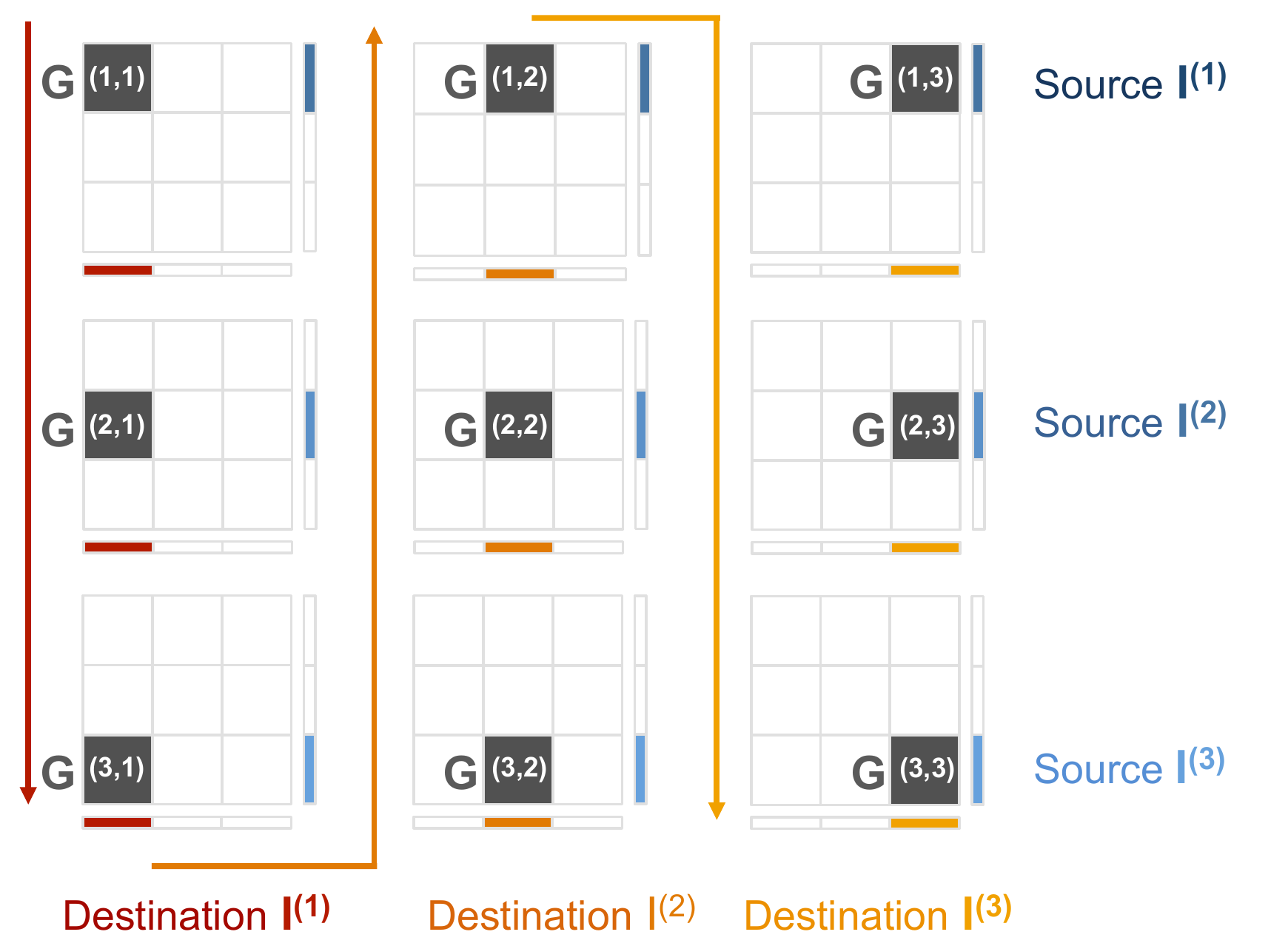}      
			\caption{
				\mflash's computation schedule for a graph with 3 intervals. Vertex intervals are represented by vertical (Source I) and horizontal (Destination I) vectors. Blocks are loaded into memory, and processed in a vertical zigzag manner, indicated by the sequence of red, orange and yellow arrows.
				This enables the reuse of input, as when going from $G^{(3,1)}$ to $G^{(3,2)}$, \mflash reuses source node interval $I^{(3)}$), which reduces data transfer from disk to memory.
			}    
			\label{fig:blockreading}
		\end{figure}
		
		\subsection{The \mflash Processing Model}\label{section:model}
		This section presents our proposed processing model. We first describe the two strategies targeted at processing dense or sparse blocks. Next, we present the novel cost-based optimization used to determine the best processing strategy.
		
		\noindent\textbf{Dense Block Processing (DBP)}:
		Figure~\ref{fig:blockreading} illustrates the DBP; notice that vertex intervals are represented by vertical (Source I) and horizontal (Destination I) vectors. After partitioning the graph into \emph{blocks}, we process them in a vertical zigzag order, as illustrated. There are three reasons for this order:
		(1) we store the computation results in the destination vertices; so, we can ``pin'' a destination interval (e.g., $I^{(1)}$) and process all the vertices that are sources to this destination interval (see the red vertical arrow); 
		(2) using this order leads to fewer reads because the attributes of the destination vertices (horizontal vectors in the illustration) only need to be read once, regardless of the number of source intervals.
		(3) after reading all the blocks in a column, we take a ``U turn'' (see the orange arrow) to benefit from the fact that the data associated with the previously-read source interval is already in memory.
		
\begin{figure}[!t]
			\centering
			\includegraphics[width=325px]{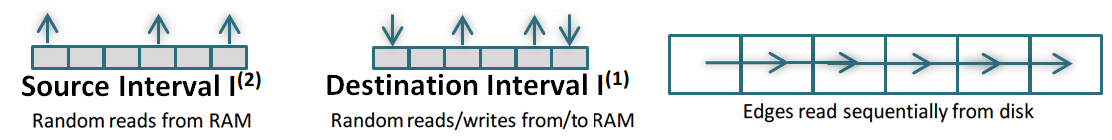}      
			\caption{Example of DBP I/O operations to process the \textit{dense} block $G^{(2,1)}$.}
			\label{fig:approach1}
		\end{figure}

		Within a block, besides loading the attributes of the source and destination intervals of vertices into RAM, the corresponding edges $e = \left \langle source, destination, edge\ properties \right \rangle$ are sequentially read from disk, as explained in Figure~\ref{fig:approach1}. These edges, then, are processed using a user-defined function so to achieve the desired computation. After all blocks in a column are processed, the updated attributes of the destination vertices are written to disk.\\
		\noindent\textbf{Streaming Partition Processing (SPP)}: 
		\noindent{The performance of DBP decreases for graphs with sparse blocks; this is because, for a given block, we have to read more data from the source intervals of vertices than from the very blocks of edges. In such cases, SPP processes the graph using partitions instead of blocks. A graph \emph{partition} is a set of \emph{blocks} sharing the same \emph{source node interval} -- a line in the logical partitioning, or, similarly, a set of \emph{blocks} sharing the same \emph{destination node interval} -- a column in the logical partitioning. Formally, a \emph{source-partition $SP^{(p)} = \bigcup _{q} G^{(p, q)}$} contains all the blocks with edges having source vertices in the interval $I^{(p)}$; a \emph{destination-partition $DP^{(q)} = \bigcup _{p} G^{(p, q)}$} contains all the blocks with edges having destination vertices in the interval $I^{(q)}$. For example, in Figure \ref{fig:blocks}, $SP^{(1)}$ is the union of blocks $G^{(1, 1)}$, $G^{(1, 2)}$, and $G^{(1, 3)}$; meanwhile, $DP^{(3)}$ is the union of blocks $G^{(1, 3)}$, $G^{(2, 3)}$, and $G^{(3, 3)}$. In a graph, hence, there are $\beta$ \emph{source-partitions} and $\beta$ \emph{destination-partitions}.}
		
		\begin{figure}[!t]
			\centering
			\includegraphics[width=325px]{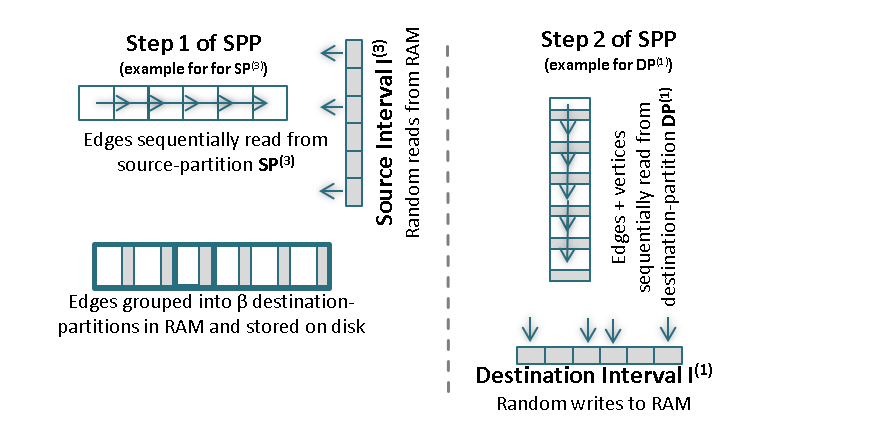}
			\caption{I/O operations for SPP taking $SP^{(3)}$ and $DP^{(1)}$ as ilustrative examples. Step 1: the edges of \emph{source-partition} $SP^{(3)}$ are sequentially read and combined with the values of their source vertices from $I_{(3)}$. Next, edges are grouped by destination, and written to $\beta$ files, one for each \emph{destination partition}. Step 2: the files corresponding to \emph{destination-partition} $DP^{(1)}$ are sequentially processed according to a given desired computation, with results written to destination vertices in $I_{(1)}$.}
			\label{fig:approach2}
		\end{figure}
		Considering the graph organized into partitions, SPP takes two steps (see Figure~\ref{fig:approach2}). In the first step, for a given \emph{source-partition} $SP^{(p)}$, it loads the values of the vertices of the corresponding interval $I^{(p)}$; next, it reads the edges of the partition $SP^{(p)}$ sequentially from disk, storing them in a buffer together with their source-vertex values. At this point, it sorts the buffer in memory, grouping the edges by destination. Finally, it stores the edges on disk into $\beta$ files, one for each of the $\beta$ \emph{destination-partitions}. This processing is performed for each \emph{source-partition} $SP^{(p)}$, $1 \leq p \leq \beta$, so to iteratively build the $\beta$ \emph{destination-partitions}.
				
		In the second step, after processing the $\beta$ \emph{source-partitions} (each with $\beta$ blocks), it is possible to read the $\beta$ files according to their destinations, so to logically ``build'' $\beta$ \emph{destination-partitions} $DP^{(q)}$, $1 \leq q \leq \beta$, each one containing edges together with their source-vertex values.    
		For each \emph{destination-partition} $DP^{(q)}$, we read the vertices of interval $I^{(q)}$; next, we sequentially read the edges, processing their values through a user-defined function. This function uses the properties of the vertices and of the edges to perform specific computations whose results will update the vertices. Finally, SPP stores the updated vertices of interval $I^{(q)}$ back on disk.\\
		
		\noindent{\textbf{Bimodal Block Processing (BBP)}}: 
		Schemes \emph{DBP} and \emph{SPP} improve the performance in complementary circumstances. But, \textit{How can we decide which processing scheme to use when we are given a graph block to process?} To answer this question, we join DBP and SPP into a single scheme -- the Bimodal Block Processing (BBP). The combined scheme uses the theoretical I/O cost model proposed by Aggarwal and Vitter \cite{Aggarwal:1988:ICS:48529.48535} to decide for \emph{SBP} or \emph{SPP}. In this model, the I/O cost of an algorithm is equal to the number of \emph{disk blocks} with size $B$ transferred between disk and memory plus the number of non-sequential reads (seeks). Since we use this model to choose the scheme with the smaller cost, we need to algebraically determine the cost of each scheme, as follows.
		
		For processing a graph $G=\{V,E\}$, \emph{DBP} performs the following operations: it reads the $|V|$ vertices $\beta$ times and it writes the $|V|$ vertices once; it also reads the $|E|$ edges once -- \emph{disk blocks} of size $B$, vertices and edges with constant sizes omitted from the equation for simplification. $\beta^2$ seeks are necessary because the edges are read sequentially. Hence, the I/O cost for \emph{DBP} is given by:
		{\small
			\begin{equation}
				\begin{split}
					\mathcal{O}\left (\mbox{DBP}\left ( G  \right ) \right ) &= \mathcal{O} \left ( \frac{(  \beta + 1) \left | V \right | + \left | E \right |  }{B} + \beta^2 \right )
				\end{split}
				\label{eq:DBP}
			\end{equation}
		}
		In turn, \emph{SPP} initially reads the $|V|$ source vertices and the $|E|$ edges; then, still in its first step, it sorts (simple shuffling) the $|E|$ edges grouping them by destination into a set of edges and vertices $|\hat{E}|$, writing them to disk. In its second step, it reads the $\hat{E}$ edges/vertices to perform the update operation, writing $|V|$ destination vertices back to disk. The I/O cost for \emph{SPP} comes to be:
		{\small
			\begin{equation}
				\begin{split}
					\mathcal{O}\left (\mbox{SPP}\left ( G  \right ) \right ) &= \mathcal{O} \left ( \frac{2 \left | V \right | + \left | E \right | + 2 \left | \hat{E} \right | }{B} + \beta \right )
				\end{split}
				\label{eq:SPP}
			\end{equation}
		}
		Equations \ref{eq:DBP} and \ref{eq:SPP} define the I/O cost for one processing iteration over the whole graph $G$. However, in order to decide in relation to the graph blocks, we are interested in the costs of Equations \ref{eq:DBP} and \ref{eq:SPP} divided by the number of graph blocks $\beta^2$. The result, after the appropriate algebra, reduces to Equations \ref{eq:DBPb} and \ref{eq:SPPb}.
		{\small
			\begin{align}
				\mathcal{O}\left (\mbox{DBP}\left ( G^{(p,q)}  \right ) \right ) &= \mathcal{O} \left ( \frac{ \vartheta \phi \left( 1 + 1/\beta \right ) +  \xi \psi }{B} \right )
				\label{eq:DBPb} \\
				\mathcal{O}\left (\mbox{SPP}\left ( G^{(p,q)}  \right ) \right ) &= \mathcal{O} \left ( \frac{ 2 \vartheta \phi /\beta + 2 \xi (\phi + \psi)  + \xi \psi }{B} \right )
				\label{eq:SPPb}
			\end{align}}
			\noindent{where $\xi$ is the number of edges in $G^{(p,q)}$, $\vartheta$ is the number of vertices in the interval, and $\phi$ and $\psi$ are, respectively, the number of bytes to represent a vertex and an edge $e$. Once we have the costs per graph block of DBP and SPP, we can decide between one and the other by simply analyzing the ratio SPP/DBP:}\\
			{\small
				\begin{align}
					\mathcal{O} \left(\frac{\mbox{SPP}}{\mbox{DBP}} \right ) &= \mathcal{O} \left ( \frac{1}{\beta} + \frac{2 \xi}{\vartheta} \left [ 1 + \frac{\psi}{\phi} \right] \right )
					\label{eq:ratio}
				\end{align}
			}
			\noindent{This ratio leads to the final decision equation:}
			{\small
				\begin{align}
					\mbox{BlockType}\left ( G^{(p,q)}  \right ) &=
					\left\{
					\begin{matrix} \mbox{sparse,} &\ if\ \mathcal{O} \left(\frac{\mbox{SPP}}{\mbox{DBP}} \right ) <1 \\ \mbox{dense,}  & \mbox{otherwise}\end{matrix} \right.
					\label{eq:blocktype}
				\end{align}
			}
			We apply Equation \ref{eq:ratio} to select the best option according to Equation \ref{eq:blocktype}. With this scheme, BBP is able to select the best processing scheme for each graph block. In Section \ref{sec:evaluation}, we demonstrate that this procedure yields a performance superior than the current state-of-the-art frameworks.
			
			\subsection{Programming Model in \mflash}\label{section:pmodel}
			The \mflash's computational model, which we named \textit{MAlgorithm} (short for \textit{Matrix Algorithm Interface}) is shown in Algorithm \ref{alg:malgorithm}.
			Since \textit{MAlgorithm} is a vertex-centric model, it stores computation results in the destination vertices, allowing for a vast set of iterative computations, such as PageRank, Random Walk with Restart, Weakly Connected Components, Sparse Matrix Vector Multiplication, Eigensolver, and Diameter Estimation.
			
			\begin{algorithm}[!t]
				\footnotesize
				\caption{\textit{MAlgorithm}: Algorithm Interface for coding in \mflash}
				\label{alg:malgorithm}
				\begin{algorithmic}
					\State \emph{\textbf{initialize}} (Vertex v)
					\State \emph{\textbf{process}} (Vertex u, Vertex v, EdgeData data)
					\State \emph{\textbf{gather}} (Accum v\_1, Accum v\_2, Accum v\_out)
					\State \emph{\textbf{apply}} (Vertex v)
				\end{algorithmic}
			\end{algorithm}
			
			The \textit{MAlgorithm} interface has four operations: 
			\textbf{initialize}, \textbf{process}, \textbf{gather}, and \textbf{apply}. 
			The \emph{initialize} operation, optionally, loads the initial value of each destination vertex; 
			the \emph{process} operation receives and processes the data from incoming edges (neighbors) -- this is where the desired processing occurs; 
			the \emph{gather} operation joins the results from the multiple threads so to consolidate a single result;
			finally, the \emph{apply} operation is able to perform finalizing operations, such as normalization -- apply is optional.
			
			\begin{algorithm}[!t]
				\footnotesize
				\caption{PageRank in \mflash}
				\label{alg:pagerank}
				\begin{algorithmic}
					\State \emph{\textbf{degree(v):}} = out degree for Vertex v
					\State \emph{\textbf{initialize(v)}}: v.value = 0
					\State \emph{\textbf{process(u, v, data)}}: v.value +=  u.value/ \emph{degree(u)}
					\State \emph{\textbf{gather(v\_1, v\_2, v\_out)}}: v\_out = v\_1 + v\_2
					\State \emph{\textbf{apply(v)}}: v.value = 0.15 + 0.85 * v.value
				\end{algorithmic}
			\end{algorithm}
			
			\subsection{System Design \& Implementation}\label{section:sdesign}
			\mflash starts by preprocessing an input graph dividing the edges into $\beta$ partitions and counting the number of edges per logical block ($\beta^2$ blocks), at the same time that the blocks are classified as sparse or dense using Equation \ref{eq:blocktype}. Note that \mflash does {\it \underline{not}} sort the edges during preprocessing, it simply divides them into $\beta^2$ blocks, $\beta^2 \ll \left | V \right |$. In a second preprocessing, \mflash processes the graph according to the organization given by the concept of \emph{source-partition} as seen in Section \ref{section:model}. At this point, blocks are only a logical organization, while partitions are physical. The \emph{source-partitions} are read and, whenever a dense block is found, the corresponding edges are extracted from the partition and a file is created for this block in preparation for DBP; the remaining edges in the \emph{source-partition} will be ready for processing using SPP. Notice that, after the second preprocessing, the logical blocks classified as dense, are materialized into physical files. The total I/O cost for preprocessing is $\frac{4 |E|}{B}$, where $B$ is the size of each block transferred between disk and memory. Algorithm \ref{alg:mflash} shows the pseudo-code of \mflash.
			\begin{algorithm}[!t]
				\footnotesize
				\caption{Algorithm \mflash}
				\label{alg:mflash}
				\begin{algorithmic}[1]
					\Require Graph $G(V, E)$ and vertex attributes $\gamma$
					\Require user-defined \emph{MAlgorithm} program
					\Require memory size $M$ and number of iterations \emph{iter}
					\Ensure vector $v$ with vertex results
					\State set $\phi$ from $\gamma$ attributes, and $\beta$ using equation \ref{eq:beta}. $\vartheta = \left | V \right | / \beta $
					\State execute graph preprocessing and \emph{partitioning} 
					\For {$i = 1 \mbox{ to \emph{iter}}$}
					\State execute the first step of \emph{SPP} (Figure \ref{fig:approach2}) to process the sparse source-partitions
					\For {$q = 1 \mbox{ to } \beta$}
					\State load vertex values of destination interval
					$I^{(q)}$
					\State initialize $I^{(q)}$ of $v$ using \emph{MAlgorithm}.initialize
					\If {there is a sparse destination-partition associated with $I^{(q)}$}
					\State \textbf{for each} edge 
					\State \hspace{4mm} invoke \emph{MAlgorithm}.process storing results in vector $v$
					\EndIf
					\If {$q$ is odd}
					\State $\mbox{partition-order} = \{1 \mbox{ to } \beta \}$ 
					\Else
					\State $\mbox{partition-order} = \{ \beta \mbox{ to } 1\}$ 
					\EndIf
					\For {$p = \{\mbox{partition-order}\}$}
					\If {$G^{(p,q)}$ is dense }
					\State load vertex values of source interval $I^{(p)}$
					\State \textbf{for each} edge in $G^{(p,q)}$ 
					\State \hspace{4mm} invoke \emph{MAlgorithm}.process storing results in vector $v$
					\EndIf
					\EndFor
					\State invoke \emph{MAlgorithm}.gather for $I^{(q)}$ of $v$
					\State invoke \emph{MAlgorithm}.apply for $I^{(q)}$ of $v$
					\State store interval $I^{(q)}$ of vector $v$ 
					\EndFor
					\EndFor        
				\end{algorithmic}
			\end{algorithm}
			
			\section{Evaluation}\label{sec:evaluation}
			We compare \mflash (\url{https://github.com/M-Flash}) with multiple state-of-the-art approaches: GraphChi, TurboGraph, X-Stream, MMap, and GridGraph. For a fair comparison, we used the experimental setups recommended by the respective authors. GridGraph did not publish nor share its code, so the comparison is based on the results reported in its publication. We omit the comparison with GraphTwist because it is not accessible and its published results are based on a hardware that is less powerful than ours. 
			We use four graphs at different scales (See Table \ref{tab:datasets}), and we compare the runtimes of all approaches for two well-known essential algorithms PageRank (Subsection~\ref{sec:pagerank}) and Weakly Connected Components (Subsection \ref{sec:cc}).
			To demonstrate how \mflash generalizes to more algorithms, we implemented the Lanczos algorithm (with \textit{selective orthogonalization}), which is one of the most computationally efficient approaches to computing eigenvalues and eigenvectors~\cite{Kang:2009:PPG:1674659.1677058} (Subsection~\ref{sec:spectral}). To the best of our knowledge, \mflash provides the \textbf{first design and implementation} of Lanczos that can handle graphs with more than one billion nodes.
			Next, in Subsection~\ref{sec:memory}, we show that \mflash maintains its high speed even when the machine has little RAM (including extreme cases, like 4GB), in contrast to the other methods. Finally, through a theoretical analysis of I/O, we show the reasons for the performance increase using the BBP strategy (Subsection~\ref{subsec:ta}).
			
			\subsection{Experimental Setup}\label{sec:setup}
			All experiments ran on a standard personal computer equipped with a four-core Intel i7-4500U CPU (3 GHz), 16 GB RAM, and 1 TB 540-MB/s (max) SSD disk.
			Note that \mflash does \textit{\underline{not}} require an SSD to run, which is not the case for all frameworks, like TurboGraph. We used an SSD, nevertheless, to make sure that all methods can perform at their best. Table \ref{tab:datasets} shows the datasets used in our experiments. GraphChi, X-Stream, MMap, and M-Flash ran on Linux Ubuntu 14.04 (x64). TurboGraph ran on Windows (x64).
			All the reported times correspond to the average time of three \textbf{cold} runs, that is, with all caches and buffers purged between runs to avoid any potential advantage due to caching or buffering.
			
			\setlength{\tabcolsep}{5pt}
			\begin{table}[ht]
				\footnotesize
				\centering
				\caption{Graph datasets used in our experiments.}
				\scalebox{1}{
					\begin{tabular}{l r r r}
						\toprule
						
						\textbf{Graph}&\hspace{5mm}\textbf{Nodes}& \hspace{5mm}\textbf{Edges} & \hspace{5mm}\textbf{Size}\\
						
						\midrule
						
						\textbf{LiveJournal \cite{Backstrom:2006:GFL:1150402.1150412}}   & 4,847,571     & 68,993,773   & Small \\
						\textbf{Twitter \cite{Kwak:2010:TSN:1772690.1772751}}       & 41,652,230    & 1,468,365,182 & Medium\\
						\textbf{YahooWeb}      & 1,413,511,391 & 6,636,600,779 & Large\\
						\textbf{R-Mat (Synthetic graph)}      & 4,000,000,000 & 12,000,000,000 & Large\\
						
						\bottomrule
					\end{tabular}}
					\label{tab:datasets}
				\end{table}
				
				\begin{table}[ht]
					\footnotesize    
					\centering
					\scalebox{0.85}{
						\begin{tabular}{l r r r r r r r}
							\toprule
							
							\textbf{}& \textbf{GraphChi}& \textbf{X-Stream} & \textbf{TurboGraph} & \textbf{MMap} & \textbf{GridGraph} & \textbf{\mflash} \\
							
							\midrule
							\textbf{PageRank} \\
							{LiveJournal (10 iter.)}  & 33.1 & 10.5 & 7.9 & 18.2 & 6.4 & \textbf{\underline{5.3}}\\
							{Twitter (10 iter.)}   & 1,199 & 962 & 241& 186& 269& \textbf{\underline{173}}\\
							{YahooWeb (1 iter.)} & 642 & 668& 628& 1,245& 235.95& \textbf{\underline{195}}\\
							{R-Mat (1 iter.)} & 2,145 & 1,360& - & - &  -& \textbf{\underline{745}}\\
							
							\midrule
							\multicolumn{5}{l}{\textbf{Connected Components}} \\
							{LiveJournal (Union Find)}   & 3.2 & 5.7& 4.4& 10.7& 4.4& \textbf{\underline{1.3}}\\
							{Twitter (Union Find)}  & 70 & 1,038& 128& 45 & 287& \textbf{\underline{25}}\\
							{YahooWeb (WCC - 1 iter.)} & 668 & 889& -& -& -& \textbf{\underline{125}}\\
							{R-Mat (WCC - 1 iter.)} & 3,334 & 2,167.63& -&- & -& \textbf{\underline{663.17}}\\
							\bottomrule
						\end{tabular}
					}
					\caption{Runtime (in seconds) with 8GB of RAM. The symbol ``-'' indicates that the corresponding system failed to process the graph or the information is not available in the respective papers.}
					\label{tab:processing}
				\end{table}
				
				\subsection{PageRank}\label{sec:pagerank}
				Table \ref{tab:processing} presents the PageRank runtime of all the methods, as discussed next.
				
				\textbf{LiveJournal} (small graph): 
				Since the whole graph fits in RAM, all approaches finish in seconds. Still, \mflash was the fastest, up to 6X faster than GraphChi, 3X than MMap, and 2X than X-Stream.
				
				\textbf{Twitter} (medium graph):
				The edges of this graph do not fit in RAM (it requires 11.3GB) but its node vectors do.
				\mflash had a similar performance if compared to MMap, however, MMap is not a generic framework, rather it is based on dedicated implementations, one for each algorithm. Still, \mflash was faster.
				In comparison to GraphChi and X-Stream, the related works that offer generic programming models, \mflash was the fastest, 5.5X and 7X faster, respectively.
				
				\textbf{YahooWeb} (large graph):
				For this billion-node graph, neither its edges nor its node vectors fit in RAM; this challenging situation is where \mflash has notably outperformed the other methods. The results of table \ref{tab:processing} confirm this claim, showing that \mflash provides a speed that is 3X to 6.3X faster that those of the other approaches.
				
				\textbf{R-Mat (Synthetic large graph)}: For our big graph, we compared only GraphChi, X-Stream, and \mflash because TurboGraph and MMap require indexes or auxiliary files that exceed our current disk capacity. GridGraph was not considered in the comparison because its paper does not provide information about R-Mat graphs with a similar scale. Table \ref{tab:processing} shows that \mflash is 2X and 3X faster that X-Stream and GraphChi respectively.
				
				\subsection{Weakly Connected Components (WCC)}\label{sec:cc}			
				When there is enough memory to store all the vertex data, the \textit{Union Find} algorithm \cite{Tarjan:1984:WAS:62.2160} is the best option to find all the WCCs in one single iteration. Otherwise, with memory limitations, an iterative algorithm produces identical solutions. Hence, in this round of experiments, we use Algorithm \textit{Union Find} to solve WCC for the small and medium graphs, whose vertices fit in memory; and we use an iterative algorithm for the YahooWeb graph.
				
				Table \ref{tab:processing} shows the runtimes for the LiveJournal and Twitter graphs with 8GB RAM;
				all approaches use Union Find, except X-Stream. This is because of the way that X-Stream is implemented, which handles only iterative algorithms.
				
				In the WCC problem, \mflash is again the fastest method with respect to the entire experiment:
				for the LiveJournal graph, \mflash is 3x faster than GraphChi, 4.3X than X-Stream, 3.3X than TurboGraph, and 8.2X than MMap. For the Twitter graph, \mflash's speed is 2.8X faster than GraphChi, 41X than X-Stream, 5X than TurboGraph, 2X than MMap, and 11.5X than GridGraph.
				
				In the results of the YahooWeb graph, one can see that \mflash was significantly faster than GraphChi, and X-Stream. Similarly to the PageRank results, \mflash is pronouncedly faster: 5.3X faster than GraphChi, and 7.1X than X-Stream. 				
				
				\subsection{Spectral Analysis using the Lanczos Algorithm}\label{sec:spectral}
				Eigenvalues and eigenvectors are at the heart of numerous algorithms, such as singular value decomposition (SVD) \cite{Berry01041992}, spectral clustering, triangle counting \cite{Tsourakakis:2008:FCT:1510528.1511415}, and tensor decomposition \cite{doi:10.1137/07070111X}. Hence, due to its importance, we demonstrate \mflash over the \textit{Lanczos algorithm}, a state-of-the-art method for eigen computation. We implemented it using method {\it Selective Orthogonalization} (\emph{LSO}). To the best of our knowledge, \mflash provides the \textbf{first design and implementation} that can handle Lanczos for graphs with more than one billion nodes. Different from the competing works, \mflash provides functions for basic vector operations using secondary memory. 
				Therefore, for the YahooWeb graph, we are not able to compare it with the other  frameworks using only 8GB of memory.
				
				To compute the top 20 eigenvectors and eigenvalues of the YahooWeb graph, one iteration of \emph{LSO} over \mflash takes 737s when using 8GB of RAM. For a comparative panorama, to the best of our knowledge, the closest comparable result of this computation comes from the HEigen system~\cite{10.1109/TKDE.2012.244}, at 150s for one iteration; note however that, it was for a much smaller graph with 282 million edges (23X fewer edges), using a {\it \underline {70-machine}} Hadoop cluster, while our experiment with \mflash used a single personal computer and a much larger graph.
								
				\subsection{Effect of Memory Size}\label{sec:memory}
\begin{figure}[t]
					\centering
					\includegraphics[width=300px]{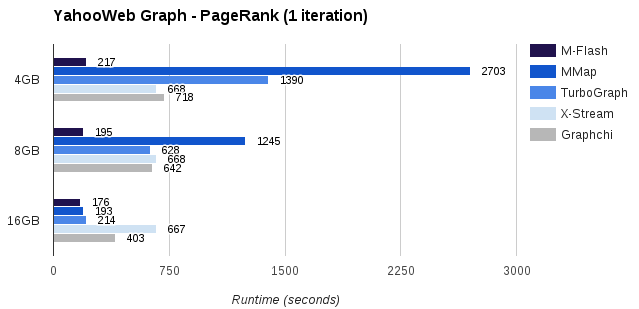}      
					\caption{Runtime comparison for PageRank over the YahooWeb graph. \mflash is significantly faster than all the state-of-the-art competitors for three different memory settings, 4GB, 8GB, and 16GB.}
					\label{fig:scalability}
				\end{figure}

				Since the amount of memory strongly affects the computation speed of single-node graph processing frameworks, here, we study the effect of memory size.
				Figure \ref{fig:scalability} summarizes how all approaches perform under 4GB, 8GB, and 16GB of RAM when running one iteration of PageRank over the YahooWeb graph.
				\mflash continues to run at the highest speed even when the machine has very little RAM, 4GB in this case. Other methods tend to slow down. In special, MMap does not perform well due to \textit{thrashing}, a situation when the machine spends a lot of time on mapping disk-resident data to RAM or unmapping data from RAM, slowing down the overall computation. For 8GB and 16GB, respectively, \mflash outperforms all the competitors for the most challenging graph, the YahooWeb. Notice that all the methods, but for \mflash and X-Stream, are strongly influenced by restrictions in memory size; according to our analyses, this is due to the higher number of data transfers needed by the other methods when not all the data fit in the memory. Despite that X-Stream worked efficiently for any memory setting, it still has worse performance if compared to \mflash because it demands three full disk scans in every case -- actually, the innovations of \mflash, as presented in Section \ref{sec:mflash}, were designed to overcome such problem.
				
				\subsection{Theoretical (I/O) Analysis}\label{subsec:ta}
				Following, we show the theoretical scalability of \mflash when we reduce the available memory at the same time that we demonstrate why the performance of \mflash improves when we combine DBP and SPP into BBP, instead of using DBP or SPP alone. Here, we use a measure that we named \emph{t-cost}; 1 unit of t-cost corresponds to three operations, one reading of the vertices, one writing of the vertices, and one reading of the edges. In terms of computational complexity, t-cost is defined as follows:
				{\small
					\begin{equation}
						\mbox{t-cost}(G(E, V)) = 2  \left | V \right | + \left | E \right |  
					\end{equation}
				}
				Notice that this cost considers that reading and writing the vertices have the same cost; this is because the evaluation is given in terms of computational complexity. For more details, please refer to the work of McSherry {\it et al.} \cite{mcsherry2015scalability}, who draws the basis of this kind of analysis.
				
				We measure the t-cost metric to analyze the theoretical scalability for processing schemes $DBP$ only, $SPP$ only, and $BBP$. 
				We perform these analyses using MatLab simulations that were validated empirically. We considered the characteristics of the three datasets used so far, LiveJournal, Twitter, and YahooWeb. For each case, we calculated the t-cost (y-axis) as a function of the available memory (x-axis), which, as we have seen, is the main constraint for graph processing frameworks.
				
\begin{figure}[!t]
					\centering
					\includegraphics[width=250px]{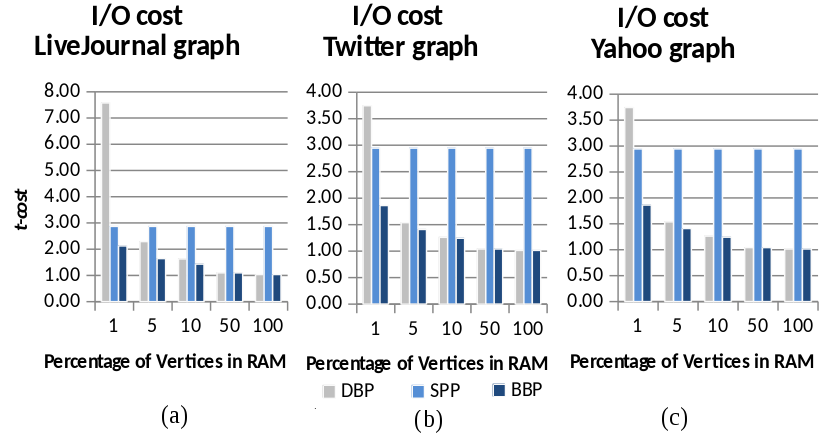}      
					\caption{I/O cost using \emph{DBP}, \emph{SPP}, and \emph{BBP} for LiveJournal, Twitter and YahooWeb Graphs using different memory sizes. \emph{BBP} model always performs fewer I/O operations on disk for all memory configurations.}
					\label{fig:graphram}
				\end{figure}

\begin{figure}[!t]
					\centering
					\includegraphics[width=280px]{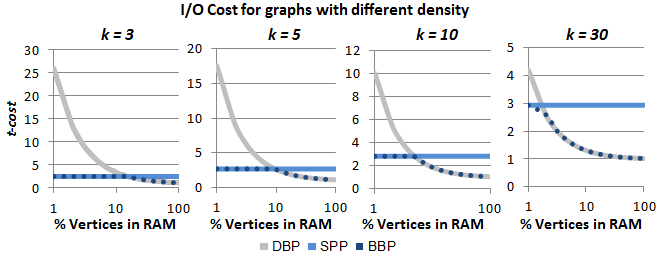}      
					\caption{I/O cost using DBP, SPP, and BBP for a graph with average degree (density) $k = \{3, 5, 10, 30 \}$, where $ |E| \approx k |V| $, and varying amount of memory}
					\label{fig:density}
				\end{figure}

				Figure \ref{fig:graphram} shows that, for all the graphs, DBP-only processing is the least efficient when memory is reduced; however, when we combine DBP (for dense region processing) and SPP (for sparse region processing) into BBP, we benefit from the best of both worlds. The result corresponds to the best performance, as seen in the charts. Figure \ref{fig:density} shows the same simulated analysis -- t-cost (y-axis) in function of the available memory (x-axis), but now with an extra variable: the density of hypothetical graphs, which is assumed to be uniform in each analysis. Each plot, from (a) to (d) considers a different density in terms of average vertex degree, respectively, 3, 5, 10, and 30. In each plot, there are two curves, one corresponding to DBP-only, and one for SPP-only; and, in dark blue, we depict the behavior of \mflash according to combination BBP. Notice that, as the amount of memory increases, so does the performance of DBP, which takes less and less time to process the whole graph (decreasing curve). SPP, in turn, has a steady performance, as it is not affected by the amount of memory (light blue line). In dark blue, one can see the performance of BBP; that is, which kind of processing will be chosen by Equation \ref{eq:blocktype} at each circumstance. For sparse graphs, Figures \ref{fig:density}(a) and \ref{fig:density}(b), SPP answers for the greater amount of processing; while the opposite is observed in denser graphs, Figures \ref{fig:density}(c) and \ref{fig:density}(d), when DBP defines almost the entire dark blue line of the plot.
				
				These results show that the graph processing must take into account the density of the graph at each moment (block) so to choose the best strategy. It also explains why \mflash improves the state of the art. 
				It is {\it important} to note that no former algorithm considered the fact that most graphs present varying density of edges (dense regions with many more edges than other regions that are sparse). Ignoring this fact leads to a decreased performance in the form of a higher number of data transfers between memory and disk, as we empirically verified in the former sections.
				
				\subsection{Preprocessing Time}
				Table \ref{tab:preprocessing} shows the preprocessing times for each graph using 8GB of RAM. As one can see, \mflash has a competitive preprocessing runtime. It reads and writes two times the entire graph on disk, which is the third best performance, after MMap and X-Stream. GridGraph and GraphTwist, in turn, demand a preprocessing that divides the graph using blocks in a way similar to M-Flash. We did not compare preprocessing with these frameworks because, as already discussed, we do not have their source code. Despite the extra preprocessing time required by M-Flash -- if compared to MMap and X-Stream, the total processing time (preprocessing + \textit{ \underline{processing with only one iteration}}) for algorithms PageRank and WCC over the YahooWeb graph, is of $1,460$s and $1,390$s, still, 29\% and 4\% better than the total time of MMap and X-Stream respectively. Note that the algorithms are iterative and \mflash needs only one iteration to overcome its competitors.
				\begin{table}[ht]
					\footnotesize    
					\centering
					\caption{Preprocessing time (seconds).}
					\scalebox{1}{
						\begin{tabular}{l r r r r}
							\toprule
							
							\textbf{}& \textbf{LiveJournal}& \textbf{Twitter} & \textbf{YahooWeb} & \textbf{R-Mat}\\
							
							\midrule
							
							\textbf{GraphChi}   & 23  & 511 & 2,781 & 7,440\\
							\textbf{X-Stream}   & \textbf{\underline{5}} & \textbf{\underline{131}} & 865 & \textbf{\underline{2,553}}\\
							\textbf{TurboGraph} & 18  & 582 & 4,694 & -\\
							\textbf{MMap}       & 17  & 372 & \textbf{\underline{636}} & -\\
							\textbf{\mflash}    & 10  & 206 & 1,265 & 4,837\\
							\bottomrule
						\end{tabular}}
						\label{tab:preprocessing}
					\end{table}
					
					\section{Conclusions}
					We proposed M-Flash, a {\it single-machine}, {\it billion-scale} graph computation framework that uses a block partition model to optimize the disk I/O. 
					M-Flash uses an innovative design that takes into account the variable density of edges observed in the different blocks of a graph. Its design uses Dense Block Processing (DBP) when the block is dense, and 
					Streaming Partition Processing (SPP) when the block is sparse. In order to take advantage of both worlds, it uses the combination of DBP and SPP according to the Bimodal Block Processing (BBP) scheme, which is able to analytically determine whether a block is dense or sparse, so to trigger the appropriate processing. To date, our proposal is the first framework that considers a bimodal approach for I/O minimization, a fact that, as we demonstrated, granted \mflash the best performance compared to the state of the art (GraphChi, X-Stream, TurboGraph, MMap, and GridGraph); notably, even when memory is severely limited.
					
					The findings observed in the design of M-Flash are a step further in determining an ultimate graph processing paradigm. We expect the research in this field to consider the criterion of block density as a mandatory feature in any such framework, consistently advancing the research on high-performance processing.
					
					\section*{Acknowledgments}
					\noindent{This work received support from Brazilian agencies CNPq (grant 444985/2014-0), Fapesp (grants 2016/02557-0, 2014/21483-2), and Capes; from USA agencies NSF (grants IIS-1563816, TWC-1526254, IIS-1217559), and GRFP (grant DGE-1148903); and Korean (MSIP) agency IITP (grant R0190-15-2012).}
					
					\bibliographystyle{splncs03}

				\end{document}